\documentclass[a4paper,dvips,12pt]{article}
\usepackage{amsmath,amssymb,exscale}   
\usepackage{array,multicol}
\usepackage{afterpage,float,flafter}   
\usepackage{epsfig,rotating,pifont,fancybox}   
\usepackage{cite}
\makeatletter   
\@addtoreset{equation}{section}   
\makeatother   
   
\newcommand{\bea}{\begin{eqnarray}}   
\newcommand{\eea}{\end{eqnarray}}   
\newcommand{\be}{\begin{equation}}   
\newcommand{\ee}{\end{equation}}

\def\Mele{{{\cal M}_{11}}}

\def\Wsix{{{\cal W}_6}}
\def\Mfour{{{\cal M}_4}}

\def\mp{M_p}

\title{   
\vspace*{-0.8cm}   
\begin{flushright}   
\normalsize{      
DFPD 02/TH/08\\
hep-th/yymmnn}\\ 
\end{flushright}    
\vspace{1cm}
\Large{\sc Isocurvature Perturbations \\
in the Ekpyrotic Universe}
\vspace*{.5cm}
\author{\large
{\sc A. Notari~$^a$  and A.~Riotto~$^b$}\\ \\
\emph{~$^a$Scuola Normale Superiore, Piazza dei Cavalieri 7, 
Pisa I-56126, Italy}\\ \\
\emph{~$^b$INFN, sezione di Padova, Via Marzolo 8,
Padova I-35131, Italy}}
}
\date{}   
\begin{document}
\maketitle
\thispagestyle{empty}
\vspace*{.5cm}

\begin{abstract}\noindent
The Ekpyrotic scenario  assumes that
our  visible Universe is a boundary brane in a 
five-dimensional bulk and that 
the hot Big Bang occurs when a nearly supersymmetric 
five-brane travelling along the
fifth dimension collides with our visible brane. 
We show that the generation of isocurvature perturbations
is a generic prediction of the Ekpyrotic Universe.
This is due to the interactions in the kinetic terms
between the  brane  modulus
parametrizing the position of the five-brane in   the
bulk  and  the dilaton and volume moduli.
We show how to separate  explicitly
the adiabatic and
isorcuvature modes by performing a rotation in field space. 
Our results  indicate that adiabatic and isocurvature pertubations
might  be
cross-correlated and that curvature perturbations
might be entirely seeded by  isocurvature perturbations.

\end{abstract}
\vspace{2.cm}   
   
\begin{flushleft}   
May 2002 \\   
\end{flushleft}
\newpage

\section{\sc Introduction}
\label{introduction}

The Ekpyrotic cosmology \cite{ek1,ek2} has recently received a great 
deal of attention because it represents an alternative to
inflationary cosmology \cite{review}. It addresses the cosmological
horizon, flatness and monopole problems and generates  a spectrum of
density
perturbations without invoking any superluminal expansion. Being 
realized in the context of eleven-dimensional heterotic M-theory, the 
Ekpyrotic scenario
is also based on   strong particle physics grounds. It assumes that
 our  visible Universe is a boundary brane in the 
five-dimensional bulk obtained 
compactifying six of the eleven dimensions on a Calabi-Yau manifold.
The hot Big Bang occurs when a nearly supersymmetric 
five-brane travelling along the
fifth dimension collides with our visible brane. 
The five-brane is attracted towards
the boundary brane where
we live by an inter-brane potential which is exponentially suppressed
when the two branes are far apart. The dynamics of
the system is described on the four-dimensional point of view
in terms of a scalar field $\varphi$, the brane modulus, 
parametrizing the separation between the two branes. 

The branes are assumed to
start widely separated  almost at rest and the four-dimensional observer
experiences a contraction of the cosmological scale factor characterized
by a 
singularity at the time when the two branes hit each other. Since the
Hubble
radius contracts faster than comoving scales, microscopic sub-horizon
 fluctuations
during the contracting phase may well 
produce curvature perturbations on cosmological
scales today \cite{ek1,ek3}. The spectrum of these adiabatic
perturbations was claimed to be  scale-independent if the brane modulus
potential scales as $e^{-c\,\frac{\varphi}{\mp}}$ 
with $c\gg 1$. This finding has been recently
challenged in a series of papers \cite{c1,c2,c3,c4,c5,c6}. 
The difficulty in determining the final spectrum
of perturbations arises from the fact that there is no prescription on 
how to match the perturbations generated in a contracting 
phase  across the  `bounce' to those in the  expanding hot big bang phase
when
radiation dominates. For
instance, if the matching 
is performed at  $\rho + \delta \rho=$ const. hypersurfaces,  
the Ekpyrotic scenario
leads to a blue spectrum with spectral index for scalar perturbations
$n_s=3$.
However, if the matching hypersurface is chosen to be  different from
 $\rho + \delta \rho=$ const., one may obtain a flat spectrum with
spectral index $n_s=1$ 
\cite{durrer}. This procedure might be motivated by the fact that  
a nonzero surface tension -- provided by some
high-energy theory ingredient -- is needed to go through the bounce,
impliying 
that the transition surface needs not to be a constant energy surface 
\cite{durrer}. 

It is fair to say that the final word about 
the power spectrum of cosmological
perturbations generated in the Ekpyrotic scenario  will be given only when 
a pure five-dimensional description of the final stages before
the brane collision is available.
In the five-dimensional picture the singularity 
is not present meaning that 
new dynamical degrees of freedom have to become relevant.
We will come back to this point in the last section.

In this paper we wish to take a  modest step and 
observe  that in the Ekpyrotic scenario
both adiabatic and isocurvature perturbations may be 
generated  during the contraction phase when the five-brane 
slowly approaches our visible world.  
A cross-correlation between entropy and curvature
perturbations 
may left imprinted. We will also  
show that curvature  perturbations may be entirely sourced by
isocurvature perturbations, thus
providing a way to produce a scale-invariant spectrum of
adiabatic perturbations, at least before the bounce. 

The paper is organized as follows. In section 2 we briefly summarize
how the effective action for the five-brane modulus is derived emphasizing
the coupling between the five-brane modulus and the dilaton and the 
volume modulus in the kinetic term. This coupling is responsible
for the generation of entropy modes. In section 3 
we describe the generation of adiabatic and isocurvature perturbations 
using the powerful technique of rotation in field space and
obtaining the equation for the gravitational potential in terms
of properly defined adiabatic and isocurvature fields. Finally, we 
end with some concluding remarks in section 4.

\section{\sc The effective action of the five-brane modulus}

Compactification of eleven-dimensional M-theory on  a Calabi-Yau threefold 
$X_6$ times $S^1/\mathbb{Z}_2$ leads to $N=1$ 
supersymmetry in four space-time dimensions. 
Five-branes configurations preserve this supersymmetry
if their world-volume $\Wsix$ is suitably aligned: it should 
enclose four-dimensional Minkowski space $\Mfour$ and a holomorphic
two-cycle ${\cal C}_2$ in the Calabi-Yau threefold, {\it i.e.}
$\Wsix=\Mfour\times {\cal C}_2$
\cite{BeckerBS, BVS, WITTENSTRONG}.
With this embedding,
the  excitations of the five-brane 
are described by a $D=6$ tensor supermultiplet
\cite{GT, KM}. The fields are 
a chiral antisymmetric tensor $B_{mn}$ $(m,n=0,1,\cdots,5)$,  
five scalar fields $X_{(1)}, \dots, X_{(5)}$ specifying 
the position of the world-volume $\Wsix$ in the eleven-dimensional
space $\Mele$ and their fermionic
partners.  
The  antisymmetric tensor
has  a self-dual field strength. This symmetry has important 
implications in four dimensions since   the effective theory will
have  a 
massless antisymmetric tensor dual to a pseudoscalar or, in terms
of supermultiplets a chiral multiplet dual to a linear 
multiplet.

Some of the $D=6$ tensor modes 
are deeply related to the Calabi-Yau geometry.
There are -- however -- universal
modes which can be more easily described, the most obvious example 
being the real scalar $ X(x^\mu)$ $(\mu=0,1,\cdots,3)$ 
associated to the position of the five-brane on the
orbifold $S^1/\mathbb{Z}_2$. 
Upon reduction to four-dimensions, the bosonic five-brane excitations
which remain are 
$B_{\mu\nu}(x^\mu)$, $B_{45} (x^\mu)$ and $X(x^\mu)$ where the 
self-duality condition 
inherited from the $D=6$ tensor relates 
$B_{\mu\nu}$ and $B_{45}$. Therefore, 
each five-brane generates  in $\Mfour$ two bosonic degrees of 
freedom. By $N=1$ supersymmetry, they belong to a 
linear multiplet (or a chiral multiplet by 
the so-called chiral-linear duality).

To describe the dynamics of the five-brane modulus 
$X$ parametrizing the position
of the five-brane in $S^1/\mathbb{Z}_2$ 
and its  couplings to four-dimensional supergravity,
one has  to couple the 
linear multiplet containing the field $X$ 
to four-dimensional supergravity. To do so, 
one writes the action  for the five-brane coupled to eleven-dimensional 
supergravity 
using the Pasti-Sorokin-Tonin  formalism to write covariant 
Lagrangians for self-dual 
(or anti-self-dual) tensors  \cite{PST}. Upon reduction, one can 
compute the effective supergravity 
couplings of the five-brane modulus $X$
to the $N=1$ supergravity and dilaton multiplets, 
the modulus of the Calabi-Yau volume and to the $E_6\times E_8$ gauge
fields 
and chiral matter on the boundaries \cite{LOW,derendinger,lukas}.

The dilaton and the universal volume modulus are respectively described 
in four-dimensions by two vector
multiplets $V$ and  $V_T$. Bianchi
identities in $\Mele$  constrain $V$ to be linear and $V_T=T+\overline T$
where $T$ is a chiral multiplet.

After properly rescaling the five-brane modulus 
$X=\kappa^2\,\varphi$, where
$\kappa=\mp^{-1}$ is the inverse of the
four-dimensional Planck scale,  
the part of the Lagrangian which is of interest to us is 
\cite{derendinger,lukas}

\begin{equation}
\label{kin}
{\cal L}_{\rm kin}=\left(\frac{1}{4}\,
\partial_\mu\chi\,\partial^\mu\chi+\frac{3}{4}\,
\partial_\mu\beta\,\partial^\mu\beta
+\frac{\tau}{2}\,
e^{\frac{\beta-\chi}{\mp}}\,
\partial_\mu\varphi\,\partial^\mu\varphi\right)\, ,
\end{equation}
where 
$\tau$ is a dimensionless constant parametrizing the five-brane
tension in units of $\kappa$ and 
$\chi$ and $\beta$ are the dilaton
and the volume modulus, respectively. The 
factor $e^{\frac{\beta-\chi}{\mp}}$ in front of the kinetic term
of the field $\varphi$ is originated from the fact that the 
the kinetic terms are 
normalized by the world-volume induced metric \cite{derendinger}
\footnote{In the chiral version of the theory where the dilaton
field is described by the real part ${\rm Re}\,S$ of the scalar
component of the chiral multiplet $S$
and the five-brane modulus belongs to a  chiral
multiplet $\hat S$,  the kinetic term
(\ref{kin})
can be understood as coming from the K\"ahler potential
$K=-{\rm log}\left(S+\overline S-\frac{\tau}{16}\frac{(\hat S+
\overline{\hat S})^2}{T+\overline T}\right)
-3\,{\rm log}\left(T+\overline
T\right)$.}.

The form (\ref{kin}) of the kinetic term of the  modulus $\varphi$
parametrizing the position of the five-brane in   
$S^1/\mathbb{Z}_2$ makes it explicit that in the
Ekpyrotic scenario the moving five-brane  induces
a nontrivial dynamics of the dilaton and volume moduli since solutions
with exactly constant $\beta$ and $\chi$ do not exist.
Similarly, 
the perturbations of the five-brane modulus, of the dilaton and of the
volume modulus might mix 
as soon as the
five-brane moves. This implies not only that both adiabatic and   
isocurvature perturbations may be created, but also that they might  be
cross-correlated and that isocurvature perturbations
may seed curvature perturbations on super-horizon scales.

\section{\sc Cosmological Perturbations}

In order to simplify the treatment of the cosmological perturbations,
from now on we consider the evolution of 
two scalar fields, the brane
modulus $\varphi$ and the dilaton $\chi$. This amounts to saying that
the volume modulus $\beta$ is tightly 
fixed at the minimum of its potential
$V(\beta)$, that is  its mass squared $\partial^2 V/\partial\beta^2$ is
much 
larger than the time-dependent term $e^{\frac{\beta-\chi}{\mp}}
\,\dot{\varphi}^2/\mp^2$. 
Needless to say, one might equivalently consider   the case in which
the dilaton $\chi$ is fixed at its minimum and the volume modulus
$\beta$ is free to move. Our following analysis applies to both
cases.

By shifting the dilaton field $\chi\rightarrow
\sqrt{2}\,\chi$ and the brane modulus $\varphi\rightarrow
\frac{e^{-\frac{\langle \beta\rangle}{2\mp}}}{\sqrt{\tau}}
\,\varphi$, 
we may write  the Lagrangian of the system as

\begin{equation}
\label{l}
{\cal L}=\frac{1}{2}\partial_\mu \chi\, \partial^\mu \chi
            + e^{-\alpha \chi} \,
\frac{1}{2}\partial_\mu \varphi \,\partial^\mu \varphi
            -V(\varphi)-V(\chi)\, ,
\end{equation}
where $\alpha= \sqrt{2}/{\mp}$,  

\begin{equation}
V(\varphi)=-V_0\, e^{-\sqrt{\frac{2}{p}}\frac{\varphi}{\mp}}\, 
\,\,\,\,(p\ll 1)
\end{equation}
is the potential of the brane modulus in the 
low energy effective action of the Ekpyrotic scenario~\cite{ek2,ek3}
and $V(\chi)$ is the dilaton potential. 
Since the field  $\varphi$ is related to the brane
separation~\cite{ek2}, at  early times when the five-brane and the
boundary brane  are
separated by a large distance the scalar field $\varphi$ is very 
large and positive and the two branes 
approach each other very slowly.

Varying the Lagrangian (\ref{l}) we obtain the equations of motion for the
homogeneous fields $\chi(t)$ and $\varphi(t)$
\begin{eqnarray}
    \ddot{\chi}+3H\dot{\chi}+\frac{\alpha}{2}\,\dot{\varphi}^2\,
        e^{-\alpha \chi}+V_\chi=0\, ,
\label{one}\\
    \ddot{\varphi}+3H\dot{\varphi}-\alpha \dot{\varphi}
        \dot{\chi} +e^{\alpha \chi}\, V_\varphi=0\, ,
\label{two}
\end{eqnarray}
were $V_x$ denotes the derivative $\frac{\partial V}{\partial x}$
and the Hubble rate $H$
in a spatially flat Friedmann-Robertson-Walker (FRW) Universe is
determined by the Friedmann equation

\begin{equation}
H^2 = \left(\frac{\dot{a}}{a}\right)^2 =\frac{8\pi }{3\mp^2} \left[
\frac{1}{2}  \dot{\chi}^{2}+e^{-\alpha \chi} \,
\frac{1}{2}  \dot{\varphi}^{2}+V(\varphi)+V(\chi)
 \right]\, , 
\end{equation}
with $a(t)$ the FRW scale factor.

In order to study the evolution of linear perturbations in the scalar
fields, we make the standard splitting $\varphi(t,{\bf x})=
\varphi(t)+\delta\varphi(t,{\bf x})$ and 
$\chi(t,{\bf x})=
\chi(t)+\delta\chi(t,{\bf x})$. Consistent study of 
the linear field fluctuations 
requires that we also consider linear scalar perturbations of the
metric.  We choose the longitudinal
gauge, where the line element perturbed to first order in the scalar
metric perturbations is (in the absence of anisotropic stress
perturbations)

\begin{equation}
ds^2=(1+2\Phi)\,dt^2-a^2(t)\,\left[(1-2\Phi)\,\delta_{ij}dx^i\,
dx^j\right]\, ,
\end{equation}
where $\Phi$ is the gravitational potential ~\cite{Bardeen,MFB,KS}.

Scalar field perturbations, with comoving wavenumber $k=2\pi
a/\lambda$ for a mode with physical wavelength $\lambda$, then
obey the perturbation equations

\begin{eqnarray}
 \ddot{\delta\varphi}+3H\dot{\delta\varphi}&+&\frac{k^2}{a^2}\delta\varphi
+V_{\varphi\varphi}\,e^{\alpha \chi}\,\delta\varphi
-4\,\dot{\varphi}\,\dot{\Phi} -2\,V_{\varphi}\,e^{\alpha
\chi}\Phi\nonumber\\
&+&\alpha\left(\dot{\chi}\dot{\delta\varphi}+\dot{\varphi}\dot{\delta\chi}
\right)
-\alpha \,V_{\varphi}\,e^{\alpha \chi}\,\delta\chi
 =0\, \label{deltadist}
\end{eqnarray}
and

\begin{eqnarray}
 \ddot{\delta\chi}+3H\dot{\delta\chi}&+&\frac{k^2}{a^2}\delta\chi
 +V_{\chi\chi}\delta \chi -4\dot{\chi}\dot{\Phi}
 -2V_{\chi}\Phi\nonumber\\  &-&
  \frac{\alpha^2}{2}\dot{\varphi}^2 e^{-\alpha
\chi}\delta\chi
 +\alpha \,\dot{\varphi}\,\dot{\delta\varphi} \,e^{-\alpha \chi}
  =0 \label{deltadilaton}\, .
\end{eqnarray}
The perturbed Einstein equations  read
\begin{eqnarray}
 \ddot{\Phi}+4H\,\dot{\Phi}+(2\dot{H}+3H^2)\Phi&=&\frac{4\pi}{\mp^2} \,
 \delta p
 \label{00}\, , \\
 \dot{\Phi}+H\Phi&=&-\frac{4\pi}{\mp^2}\, 
\delta q 
\label{0i}\, ,\\
 3H\dot{\Phi}+3H^2\Phi+\frac{k^2}{a^2}\Phi&=&-\frac{4\pi}{\mp^2}\,
 \delta \rho\, ,
 \label{ii}
\end{eqnarray}
where  the total energy and momentum perturbations are given by
\begin{eqnarray}
\delta p&=&\delta\, T_i^i=\left(\dot{\chi}\,\dot{\delta\chi} +
            \dot{\varphi}\,\dot{\delta\varphi}\,e^{-\alpha \chi}
            -\frac{\alpha}{2}\,\dot{\varphi}^2\,
e^{-\alpha \chi}\,\delta\chi\right)
     \nonumber  \\
        &-&\Phi (\dot{\chi}^2+
            \dot{\varphi}^2\,e^{-\alpha \chi})
            -V_\varphi \delta \varphi   -V_\chi \delta \chi
        \label{deltap}\, ,\\
\delta\rho&=&\delta T_0^0=\left(\dot{\chi}\,\dot{\delta\chi} +
            \dot{\varphi}\,\dot{\delta\varphi}\,e^{-\alpha \chi}
            -\frac{\alpha}{2}\,\dot{\varphi}^2\,
e^{-\alpha \chi}\,\delta\chi\right)\nonumber\\
        &-&\Phi (\dot{\chi}^2+
            \dot{\varphi}^2\,e^{-\alpha \chi})
            +V_\varphi \delta \varphi  +V_\chi \delta \chi\, ,
        \label{deltarho}\\
\delta q&=&\delta T_0^i=-\left(\dot{\chi}{\delta\chi} +
            \dot{\varphi}{\delta\varphi}\,e^{-\alpha \chi}\right)\, .
        \label{deltaq}
    \end{eqnarray}
We obtain a useful relation if we take the Eq. (\ref{0i})
times $-3H$ and then sum it to  Eq.  (\ref{ii})
\begin{equation}
\frac{k^2}{a^2}\,\Phi=-\frac{4\pi}{\mp^2}\, \epsilon_m,
\label{Phiepsilon}
\end{equation}
where
\begin{equation}
\epsilon_m\equiv\delta\rho-3H\,\delta q \label{epsilon}
\end{equation}
is used to represent the total matter perturbation.

Let us now turn to
isocurvature perturbation. 
A dimensionless definition of the total entropy perturbation
(which is automatically gauge-invariant) is given by

\begin{equation}
\label{S_total}
S = H \left( {\delta p \over \dot{p}} - {\delta\rho \over
\dot\rho} \right) \,,
\end{equation}
which can be extended to define a generalised entropy perturbation
between any two matter quantities
$x$ and $y$

\begin{equation}
S_{xy} = H \left( {\delta{x}\over \dot{x}} - {\delta{y}
\over \dot{y}} \right) \, .
\end{equation}
This implies that when two scalar fields are present in the dynamics of
the system, isocurvature perturbations are generated.

In our case, inserting the expressions for the pressure and the energy
density
as well as their perturbations, we find 

\begin{eqnarray}
S&=&\frac{2}
 {3\left(2\dot{V} + 3H\left(\dot{\chi}^2+\dot{\varphi}^2 e^{-\alpha \chi}
\right)\right)(\dot{\chi}^2+\dot{\varphi}^2 e^{-\alpha
\chi})}\times\nonumber \\
&&\left\{\left(\dot{V} + 3H \left(\dot{\chi}^2+\dot{\varphi}^2 
e^{-\alpha \chi}\right)\right)\left(V_\chi \delta\chi + V_\varphi 
\delta\varphi\right)+\right.\nonumber \\
&&\left. \dot{V}\left[\left(\dot{\chi}\dot{\delta\chi} +
             \dot{\varphi}\dot{\delta\varphi}e^{-\alpha \chi}
             -\frac{\alpha}{2}\dot{\varphi}^2 e^{-\alpha \chi}\delta\chi
\right)-\Phi\left(\dot{\chi}^2+\dot{\varphi}^2 e^{-\alpha
\chi}\right)\right]
\right\}\, .
\label{source}
\end{eqnarray}

\subsection{Rotation in field space}

In order to clarify the role of adiabatic and entropy
perturbations, their evolution and their interconnection, we define
new adiabatic and entropy fields
by a rotation in field space \cite{gordon}. The ``adiabatic field'',
$\sigma$,
represents the path length along the classical trajectory, such that

\begin{equation}
\dot{\sigma}=\cos\theta\, \dot{\chi} +
            \sin\theta \, \dot{\varphi}\, e^{-\frac{\alpha \chi}{2}}\, ,
\end{equation}
where

\begin{equation}
\cos\theta = \frac{ \dot{\chi}  }
        {\sqrt{\dot{\chi}^2+\dot{\varphi}^2 \,  
e^{-\alpha\chi}}}\, , \,\,\,
\sin\theta = \frac{\dot{\varphi}\,  e^{-\frac{\alpha \chi}{2}} }
        {  \sqrt{\dot{\chi}^2+\dot{\varphi}^2 \,e^{-\alpha\chi}}}\, .
\end{equation}
The equation of motion of the field $\sigma$ is remarkably
simple
\begin{equation}
\ddot{\sigma}+3H\,\dot{\sigma}+V_\sigma=0\, ,
\end{equation}
where

\begin{equation}
V_\sigma=\cos\theta\,  V_{\chi} +
    \sin\theta\, V_{\varphi}\,e^{\frac{\alpha \chi}{2}}\, .
\end{equation}
The fluctuation $\delta \sigma$ is
the component of the two-field perturbation vector along the
direction of the background fields' evolution \cite{gordon}.
Fluctuations orthogonal to the
background classical trajectory represent non-adiabatic
perturbations, and we define the ``entropy field'' $s$ such that

\begin{equation}
\delta s =-\sin\theta\,\delta\chi+\cos\theta\,\delta\varphi\,
e^{-\frac{\alpha\chi}{2}}\, .
 \end{equation}
From this definition, it follows that $s=$constant along the
classical trajectory, and hence entropy perturbations are
automatically gauge-invariant~\cite{StewartWalker}. Perturbations
in $\delta\sigma$, with $\delta s=0$, describe adiabatic
field perturbations, and this is why we refer to $\sigma$ as the
``adiabatic field''.

The angle $\theta$ changes with time along the classical trajectory
according to the equation 
\begin{equation}
\dot\theta=-\frac{V_s}{\dot{\sigma}} +\frac{1}{2}\,\alpha \,\sin\theta
\, \dot{\sigma}\, ,
\label{thetapunto}
\end{equation}
where 
\begin{equation}
V_s=-\sin\theta\,  V_{\chi} +
    \cos\theta\, V_{\varphi}\,e^{\frac{\alpha \chi}{2}}\, .
\end{equation}

\subsection{The equations 
for the gravitational potential and the entropy field}

Armed with these definitions, we are now ready to compute the equation
of motion of the gravitational potential $\Phi$. 
First of all, we rewrite Einstein Eqs.
(\ref{00}), (\ref{0i}) and (\ref{ii}) as
\begin{eqnarray}
 \ddot{\Phi}+4H\dot{\Phi}+(\dot{H}+3H^2)\Phi&=&
  \frac{4\pi}{\mp^2} \left[\left(\dot{\chi}\,\dot{\delta\chi} +
    \dot{\varphi}\,\dot{\delta\varphi}\,e^{-\alpha \chi}
        -\frac{\alpha}{2}\dot{\varphi}^2\,e^{-\alpha
        \chi}\,\delta\chi\right)\right.\nonumber\\
        &-&\left.V_\varphi\, \delta \varphi -V_\chi\, \delta \chi\right]\,
,
      \label{iiNEW} \\
 \dot{\Phi}+ H\Phi&=&+\frac{4\pi}{\mp^2}
 \dot{\sigma}\, \delta \sigma\, ,
  \label{0iNEW}  \\
 3H\dot{\Phi}+(\dot{H}+3H^2)\Phi+\frac{k^2}{a^2}\Phi&=&\frac{4\pi}{\mp^2}
\left[-\left(\dot{\chi}\,\dot{\delta\chi} +
    \dot{\varphi}\,\dot{\delta\varphi}\,e^{-\alpha \chi}
        -\frac{\alpha}{2}\dot{\varphi}^2\,e^{-\alpha
        \chi}\delta\chi\right)\right.\nonumber\\
        &-&V_\varphi \,\delta \varphi  -V_\chi\, \delta \chi]\, ,
 \label{00NEW}
\end{eqnarray}
where we used

\begin{equation}
\dot{H}=- \frac{4\pi}{\mp^2}\,\dot{\sigma}^2=-\frac{4\pi}{\mp^2}
\left[\dot{\chi}^2+
            \dot{\varphi}^2\,e^{-\alpha \chi}\right]\, .
\end{equation}
Summing Eqs. (\ref{00NEW}) and  (\ref{iiNEW}) and  multiplying Eq.
(\ref{0iNEW}) by
$-\frac{2}{\dot{\sigma}}(\ddot{\sigma}+3H\dot{\sigma})$, we finally obtain

\begin{equation}
\ddot{\Phi}+ \left(H-2\frac{\ddot{\sigma}}{\dot{\sigma}}\right)\dot{\Phi}+
\left(2\dot{H}-2H\frac{\ddot{\sigma}}{\dot{\sigma}}\right)\Phi +
\frac{k^2}{a^2} \Phi =
- \frac{8\pi}{\mp^2}\,V_s\, \delta s  \, .
\label{equazionePhi}
\end{equation}
This equation makes manifest the   neat separation of the roles played by
the 
adiabatic and isocurvature perturbations.
Indeed, on the left-hand side only 
the adiabatic field $\sigma$ appears while  on the 
right-hand side there is a  source 
which is proportional only to the relative entropy
perturbations $\delta s$ between the brane-modulus and the dilaton field.

Eq. (\ref{equazionePhi}) shows explicitly how on large scales entropy
perturbations can source the gravitational potential in the
Ekpyrotic cosmology. When the five-brane slowly approaches the boundary
brane
and the brane modulus decreases in time, perturbations in the 
brane modulus, in the dilaton field and in the gravitational
potential are generated. 
Furthermore,  
adiabatic and isocurvature perturbations are inevitably 
correlated. These results are similar to what found in standard inflation
when two or more scalar fields are present
\cite{gordon,bmr1,bmr2,bmr3,wbmr}.

Using now $\dot{V}=V_\sigma \dot{\sigma}$, $
\delta V=V_\sigma \delta\sigma + V_s \delta s$
and

\begin{equation}
\dot{\chi}\,\dot{\delta\chi} +
            \dot{\varphi}\,\dot{\delta\varphi}\,e^{-\alpha \chi}
            -\frac{\alpha}{2}\dot{\varphi}^2\,e^{-\alpha \chi}\delta\chi=
\dot{\sigma}\,\dot{\delta\sigma}+V_s\delta s\, ,
\end{equation}         
we can write the isocurvature source $S$ in Eq. (\ref{source}) as 

\begin{equation}
S=2\frac{(V_\sigma + 3H\dot{\sigma})
(V_\sigma \delta\sigma + V_s \delta
s)+V_\sigma(\dot{\sigma}\dot{\delta\sigma}
+V_s \delta s)-\Phi V_\sigma \dot{\sigma}^2} 
 {3(2V_\sigma + 3H\dot{\sigma})\dot{\sigma}^2}\, .
\end{equation}
Note this expression is the same as in the standard case $\alpha=0$ 
\cite{gordon}.
We can rewrite the last expression  as

\begin{eqnarray}
S&=&2\frac{-\ddot{\sigma}(V_\sigma \delta\sigma + V_s 
\delta s)+V_\sigma(\dot{\sigma}\dot{\delta\sigma}+V_s \delta s)-
\Phi V_\sigma \dot{\sigma}^2}  {3(2V_\sigma + 3H\dot{\sigma})
\dot{\sigma}^2}\nonumber\\
&=& 2\frac{V_\sigma\left(\dot{\sigma}\dot{\delta\sigma}-\Phi\dot{\sigma}^2
\
 -\ddot{\sigma}\delta\sigma\right) + V_s\delta s
\left(V_\sigma-\ddot{\sigma}\right)}  {3(2V_\sigma +
3H\dot{\sigma})\dot{\sigma}^2}\, .
\end{eqnarray}
Making use of Eq. (\ref{Phiepsilon}) 
we can express $\epsilon_m$ in terms of the new fields $\delta\sigma$ and
$\delta s$

\begin{equation}
\epsilon_m=\dot{\sigma}\left(\dot{\delta\sigma}-\dot{\sigma}\Phi\right)-
\ddot{\sigma}\,\delta\sigma
+2\,V_s\,\delta s\, .
\label{epsilonNew}
\end{equation}

Finally, using Eqs. (\ref{Phiepsilon}) and  (\ref{epsilonNew}) we get

\begin{eqnarray}
S&=&2\frac{V_\sigma \epsilon_m -2V_\sigma V_s\delta s +V_s\delta s
\left(2V_\sigma+3H\dot{\sigma}\right)}  {3(2V_\sigma + 
3H\dot{\sigma})\dot{\sigma}^2}\nonumber\\
&=&-\frac{\mp^2\,V_\sigma (\frac{k^2}{a^2}\Phi)} {6\pi (2V_\sigma +
3H\dot{\sigma})
\dot{\sigma}^2}+\frac{H V_s\delta s }{(2V_\sigma +
3H\dot{\sigma})\dot{\sigma}}
\label{en}
\, .
\end{eqnarray}

Since the curvature perturbation is given by  ${\cal
R}=\Phi-\frac{H}{\dot{H}}
\left(\dot\Phi+H\Phi\right)$, Eq. (\ref{equazionePhi}) might be rewritten
as

\begin{equation}
\dot{{\cal R}}=-3H\,\frac{\dot p}{\dot \rho}\,S\, .
\label{curvature}
\end{equation}
The change in the curvature
perturbation on large scales can therefore be directly related to the 
nonadiabatic part of the pressure perturbation \cite{book}.

Clearly, there can be significant changes in the gravitational potential
on large scales and a large cross-correlation between
the adiabatic and the isocurvature modes only 
if the entropy perturbation is not suppressed. The next step is therefore
to
compute the equation of motion of the entropy field $\delta s$. The
computation
is lengthy, but straightforward. 
One finds

\begin{equation}
      \ddot{\delta s}+3H \,\dot{\delta
      s}+\left(\frac{k^2}{a^2}+3\dot{\theta}^2+V_{ss}+ \Gamma_\alpha
\right)\delta s
      =-\frac{V_s}{\dot{\sigma}^2}
      \frac{\mp^2\,k^2}{2\pi a^2}\,\Phi\, ,
\label{sss}
\end{equation}
where we have defined

\begin{equation}
V_{ss}= \cos^2\theta\,e^{\alpha\chi}\,V_{\varphi\varphi}+\sin^2\theta\,
V_{\chi\chi}
\end{equation}
and

\begin{equation}
\Gamma_\alpha=\frac{\alpha}{2}\left(
-\cos\theta\, V_\sigma +6\sin\theta\, V_s -\frac{1}{2}\sin 2\theta\,
V_\varphi \,e^{\alpha
      \chi/2}\right)-\frac{\alpha^2}{4}\,\dot{\sigma}^2 \,\cos^2\theta\,
       -\alpha^2\,\dot{\sigma}^2 \,\sin^2\theta\,.
\end{equation}
Eq. (\ref{sss}) is the same found in Ref. \cite{gordon} apart from
the new $\alpha$-dependent  terms present in $\Gamma_\alpha$.
On large scales the inhomogeneous term proportional to the gravitational
potential $\Phi$ becomes negligible and the field $\delta s$
satisfies a homogeneous second-order equation of motion
for the entropy perturbation which is decoupled from the adiabatic
modes and metric perturbations. This amounts to saying that if
the initial entropy perturbation is zero on large scales, it will
remain zero at later times as well.

Equations (\ref{equazionePhi}) and (\ref{sss}) are the key
equations which govern the evolution of the adiabatic and entropy
perturbations in the Ekpyrotic scenario before the
collision of the five-brane and the boundary brane. In principle, 
they allow us to
follow the effect on the adiabatic curvature perturbation due to the
presence of entropy perturbations before the collision and they
provide the ``initial'' conditions for the matching technique through
the bounce in a four-dimensional approach. 

The  general
super-Hubble solution of Eqs. (\ref{equazionePhi}) and  
(\ref{sss})
can  be written as
\begin{eqnarray}
\Phi &=& f_+(t) +  f_-(t) + {\cal S}(t) \, ,\\ 
\delta s &=& 
g_+(t) +  g_-(t)\, ,
\end{eqnarray}
where the functions $f_\pm$ and $g_\pm$ are the growing/decaying (or
constant)
modes
of the homogeneous equations and ${\cal S}(t)$ is a particular integral
of the inhomogenous Eq. (\ref{sss}). From Eq. 
(\ref{equazionePhi}) one finds that the amplitude of the particular
integral
${\cal S}(t)$ is correlated with the amplitude of the entropy perturbation

\begin{equation}
\langle \Phi(k)\,\delta s^*(k^\prime)\rangle=\frac{2\pi^2}{k^3}\,
{\cal C}_{\Phi\delta s}\,\delta^{(3)}\left(k-k^\prime\right)\propto
{\cal S}(t)\, .
\end{equation}
A nonvanishing cross-correlation between the adiabatic and isocurvature 
modes are expected before collision.
One can now envisage various possibilities.

If $|\dot\chi/\dot\varphi|\ll 1$ and $V(\chi)=0$, 
the dynamics of the system is close to the single field version
of the Ekpyrotic Universe \cite{ek1}, {\it i.e.} the scale factor
is given by $a(t)\sim (-t)^p$ while the brane modulus scales like
$\varphi(t)\sim\sqrt{2\,p}\,{\rm ln}(-M\,t)$, where $M^2\simeq 
(V_0/p\mp^2) $ ($p\ll 1$).
It is easy to show that $|\dot{\chi}/\dot\varphi|
=|\alpha\,\mp\,\sqrt{p/2}|\ll 1$.
The adiabatic field $\sigma$ is given by ${\rm
exp}(-\alpha\chi/2)\,\varphi$
and the solution to the homogeneous part of 
the equation for the gravitational potential
(\ref{equazionePhi}) leads to $\Phi=A(k)\frac{H}{a}+B(k)$ at 
superhorizon scales in the collapsing
phase. The $A$-growing mode has a scale-independent spectrum, $|A|^2\,k^3$
is 
$k$-independent, while the constant mode has a blue spectrum. 
One then needs to   match this solution to
the usual (approximately constant) gravitational potential in the
radiation era. If the matching from the collapsing phase
to the radiation era is performed on constant energy surfaces, 
the gravitational potential in the radiation era inherits the blue
spectrum
from the constant mode in the collapsing phase. However, 
a nonzero surface tension -- provided by some
high-energy theory ingredient -- is needed to go through
 the bounce. This implies 
that the transition surface needs not to be a constant energy surface 
\cite{durrer}. For instance, imposing the matching on a surface where its
shear vanish, one finds that the gravitational potential in the radiation
phase
inherits the flat spectrum of $A$ \cite{durrer}. On the top of that, a 
cross-correlation between adiabatic and isocurvature modes is generated
before the collapse with ${\cal C}_{\Phi\delta s}\propto
\alpha\,p\,\mp\,
k^2={\cal O}(p)\,k^2$, {\it i.e.} a blue spectrum. 

If $|\dot\chi/\dot\varphi|\gg 1$ and $V(\chi)=-M^4\,
{\rm exp}(-c\chi/\mp)$ with $M^4\gg V_0\simeq 0$, one obtains 
a sort of modified version of the pre big bang model \cite{pbb} with a 
potential for the dilaton field. The adiabatic field
$\sigma$ is identified with  the dilaton field $\chi$ and the final
spectrum
of curvature perturbations is flat if $c=\sqrt{3} $ and the
matching through the bounce is done onto a constant energy surface
\cite{finelli}\footnote{If the matching is done using the prescription
discussed in Ref. \cite{durrer}, the spectrum is flat if $c\gg 1$.}.
Even in this  a case cross-correlation between adiabatic and isocurvature
modes
is present. Yet, it is suppressed. 

A much more interesting 
situation is realized if the brane-modulus potential is very tiny, 
$V_0\simeq
0$ and $V(\chi)=-M^4\,{\rm exp}\left(-\beta\chi/\mp\right)$, with 
$M$ some mass scale.
Going to conformal time $d\tau=dt/a$ and integrating Eq. (\ref{two}), one 
finds $\varphi^\prime=C\,{\rm exp}\left(\alpha\chi\right)/a^2$, where
primes mean derivatives with respect to conformal time $\tau$ and $C$
is an integration constant. The ansatz $a(\tau)=\left(-\mp(1-q)\tau
\right)^{\frac{q}{1-q}}$ and $\chi(\tau)=A\, {\rm ln}\left(-\mp
(1-q)\tau\right)$ satisfy the equations of motion if $\alpha\,A=
2(3q-1)/(1-q)$, $\beta\,A/\mp=2/(1-q)$ and $\alpha\mp/\beta=3q-1$.
Suppose now that  the energy density of the system is 
dominated by the kinetic term of the brane-modulus. This will
be the case if, for instance, $A\ll C/\mp$. After
introducing the new variable $\delta S=a\,\delta s$, 
Eq. (\ref{sss})
reduces to

\begin{equation}
\delta S^{\prime\prime}+\left(k^2-\frac{a^{\prime\prime}}{a}-
\frac{\alpha^2}{4}\,{\varphi^\prime}^2\,e^{-\alpha\chi}\right)
\delta S=0\, .
\end{equation}
Since $C^2\simeq 2q\,\mp^4$ and $\alpha=\sqrt{2}/\mp$, one finds 

\begin{equation}
\delta S^{\prime\prime}+\left(k^2-\frac{2\,q^2}{(1-q)^2\,\tau^2}\right)
\delta S=0\, .
\end{equation}
A nearly invariant spectrum for the entropy perturbations
is obtained if $2\,q^2/(1-q)^2\simeq 2$, or $q\simeq 1/2$. This 
is a desirable output since it means that 
adiabatic perturbations are entirely sourced by entropy perurbations
inducing a flat spectrum for curvature 
perturbations with maximum cross-correlation. 
This is the Ekpyrotic version of the mechanism to  produce
curvature perturbations from isocurvature seeds
in pre big bang models \cite{cur1} and ordinary inflation models 
\cite{cur2}.

\section{\sc Concluding remarks}

Let us conclude with some comments.
The generation  of adiabatic plus isocuvarture perturbations and 
their cross-correlation we have described in this paper occur in the
Ekpyrotic scenario before the moving five-brane collapses onto
the boundary brane. Isocurvature perturbations 
might not 
survive after the bounce  if during the subsequent period of 
reheating both the brane modulus and the dilaton
decay into the same species. 
In order to have isocurvature perturbations deep in the radiation era
after
the collapse it is necessary to have at least one non-zero isocurvature 
perturbation  $S_{\alpha \beta}\equiv\delta_{\alpha}/(1+w_{\alpha})-
\delta_{\beta}/(1+w_{\beta})
\neq 0$, 
where $\delta_{\alpha}=\delta
\rho_{\alpha}/\rho_{\alpha}$ and  $w_{\alpha}=p_{\alpha}/\rho_{\alpha}$ 
(the ratio of the pressure to the energy density) for some 
components $\alpha$ and $\beta$ 
of the system. This may happen if the  fields responsible for the
isocurvature perturbations decay into radiation and cold dark matter
at different epochs. 
On the other hand, we have seen  that 
curvature perturbations may be entirely seeded by isocurvature
perturbations, thus  providing a novel
mechanism to produce a scale-independent spectrum of adiabatic
perturbations in the Ekpyrotic Universe.

All previous discussions make it clear    that 
the features of the cosmological perturbations
after the big bang as well as the 
way reheating takes place   depend strongly
on the details of the transition from the collapsing to
the expanding phase when  the five-brane is absorbed by the boundary
brane.  In this absorption process the degree of freedom represented 
by the brane modulus gets replaced by new degrees of freedom.

In  eleven-dimensional M-theory, $E_8\times E_8$ 
gauge fields with strength $F_{z_i z_j}$
living on the boundary of the eleventh dimension and 
in the six-dimensional Calabi-Yau manifold ($z_i$ with 
$i=1,2,3$ are the complex coordinates of such a manifold) satisfy
equations
of motion of the type $F_{zz}=F_{\bar z\bar z}=
g^{z\bar z}F_{z\bar z}=
0$. Gauge field 
configurations satisfying these equations are generically called
instantons
(for istance, if the Calabi-Yau manifold is a two-dimensional torus times
a four-dimensional variety $K3$, one gets the traditional 
quantum field theory instanton equations 
$F=\pm \widetilde{F}$). In non-standard compactifications of M-theory
with a certain number $N$ of five-branes, one schematically obtains
the following constraint 

\begin{equation}
N+\int F\wedge F={\rm constant}\, ,
\label{flux}
\end{equation}
that is the number of five-branes
plus the ``number'' of instantons given by $\int F\wedge F$ is
conserved. From
Eq. (\ref{flux}) one can infer  that in M-theory 
it is possible to replace one five-brane with one instanton and
viceversa. An instanton
with $\int F\wedge F=a$
can shrink to zero size, becoming a so-called small instanton with 
$F\wedge F=a\,\delta(z_i)$ and leave the boundary brane along the 
eleventh dimension under the form of a five-brane \cite{hanany}.
A unit of instanton flux is replaced by a unit of five-brane flux
still satisfying the  constraint (\ref{flux}).

This phenomenon is similar to what happens in Type IIA string 
theory where D4 branes may get emitted or absorbed by a set of D8
brane plus O8 orientifold. A crucial role for describing this
phenomenon is played by the strings stretched between the D4 brane
and the (D8 $+$ O8) system. If they are massive, {\it i.e.} if their
length is nonzero, it means that the D4 brane is in the bulk away from
the (D8 $+$ O8) system. On the contrary, if the strings are tensionless,
their length is zero and the D4 brane touches the (D8 $+$ O8) system.
The transition may
be described from the four-dimensional point of view as a Higgs mechanism.
The D4 brane can be dscribed by an $N=2$ vector multiplet
$\{W_\alpha,\phi\}$ while the strings stretched between the D4 brane
and the (D8 $+$ O8) system are described in terms of 
some hypermultiplets $Y$. They 
appear as particles on the D4 brane. Upon compactification to
four-dimensions
the Lagrangian (in the
$N=1$ supersymmetry language) can be written as 

\begin{equation}
\int d^2\theta\left(
W_\alpha^2+ Y\phi \bar Y\right)+\int d^4\theta\left(\bar\phi \phi +
\bar Y Y\right)\, .
\end{equation}
When $\langle \phi\rangle\neq 0$, the $Y$-field is massive.
This means that the strings between  the D4 brane and the 
(D8 $+$ O8) system have a nonvanishing length (or tension):
the D4 brane is in the bulk. At the transition
point $\langle Y \rangle=\langle\phi\rangle= 0$, 
tensionless (massless) strings appear in the spectrum 
and the D4 brane is absorbed by the
(D8 $+$ O8) system. This transition  gives rise to gauge field
configurations 
whose moduli space contain a number of free parameters matching the
number of the degrees of freedom before the absorption. Some
of the $Y$-fields are now interpreted as instanton moduli.

Going back to the case of the Ekpyrotic scenario, 
a crucial role is played by two-dimensional branes, membranes, stretched
between the boundary brane and the slowly approaching five-brane
\cite{stro}.
These membranes are massive when the five-brane is in the bulk
and the vacuum expectation value of the brane modulus is nonzero.
When the five-brane touches the boundary brane, these membranes
have zero length. Therefore, one might hope to describe the 
transition from the five-brane to the small instanton in terms
of a four-dimensional Higgs mechanism as done for 
Type IIA string theory. The problem is that the massless
membranes appear as tensionless strings in the five-brane world-volume
even after compactifying down to four-dimensions.
At the transition point the relevant degrees of freedom of the 
theory are therefore tensionless strings, the anti-self-dual tensors
of the five-brane, the brane modulus and some instanton gauge field
configurations.
They are the fundamental degrees of freedom which allow the
description of the  the system during the
last moments before the collision and the subsequent big bang.
In terms of these degrees of freedom 
the theory does not admit a simple and perturbative
four-dimensional description. Nevertheless, there might be
cases in which the system is tractable. 
One might hope to do some progress if the transition
starts when the five-brane is sufficiently far from the
boundary brane and the membranes -- or better
 the corresponding strings --
are sufficiently massive to admit a  description
in terms of four-dimensional scalar fields $Y$. 
Work along these lines is in progress \cite{us}.


\section*{\sc Acknowledgments} 
We thank N. Bartolo, R. Brandenberger, 
E. Copeland, F. Finelli, D. Lyth, S. Matarrese and especially A. Zaffaroni
for useful discussions.


\begin{thebibliography}{99}
%

\bibitem{ek1} J. Khoury, B.A. Ovrut, P.J. Steinhardt and N. Turok, 
    Phys. Rev. {\bf D64}, 123522 (2001).

\bibitem{ek2}  J. Khoury, B.A. Ovrut, N. Seiberg, P.J. Steinhardt and 
        N. Turok, hep-th/0108187.


\bibitem{review} D.H. Lyth and A. Riotto, Phys. Rept. {\bf 314}, 1
(1999).

\bibitem{ek3} J. Khoury, B.A. Ovrut, P.J. Steinhardt and N. Turok, 
        hep-th/0109050.

\bibitem{c1} D. Lyth, Phys. Lett. {\bf B524}, 1 (2002). 

\bibitem{c2} R. Brandenberger and F. Finelli, JHEP 0111, 056 (2001).

\bibitem{c3}J. Hwang, Phys. Rev. {\bf D65}, 063514 (2002).

\bibitem{c4}J. Hwang and H. Noh, hep-th/0203193.

\bibitem{c5} S. Tsujikawa, Phys. Lett. {\bf B526},  179 (2002).

\bibitem{c6} J.~Martin, P.~Peter, N.~Pinto Neto and D.~J.~Schwarz,
hep-th/0112128; {\it ibidem}  hep-th/0204222.



\bibitem{durrer}
R.~Durrer and F.~Vernizzi, hep-ph/0203275.








\bibitem{BeckerBS}
K. Becker, M. Becker and A. Strominger, Nucl. Phys. {\bf B456},
130 (1995).

\bibitem{BVS}
M. Bershadsky, C. Vafa and V. Sadov,  Nucl. Phys. {\bf B463}, 420 (1996).

\bibitem{WITTENSTRONG}
E. Witten,  Nucl. Phys. {\bf B471}, 135 (1996).


\bibitem{GT}
G. W. Gibbons and P. K. Townsend, Phys. Rev. Lett. {\bf 71}, 3754 (1993).

\bibitem{KM}
D. Kaplan and J. Michelson, Phys. Rev. {\bf D53}, 3474 (1996).

\bibitem{PST}
P. Pasti, D. Sorokin and M. Tonin, Phys. Lett. {\bf B398}, 41 (1997); 
I. Bandos, K. Lechner, A. Nurmagambetov, P. Pasti, D. Sorokin and 
M. Tonin, Phys. Rev. Lett. {\bf 78}.

\bibitem{LOW}
A. Lukas, B. A. Ovrut and D. Waldram, Phys. Rev. {\bf D59}, 106005 (1999).


\bibitem{derendinger} J.~P.~Derendinger and R.~Sauser,
Nucl.\ Phys.\ B {\bf 598}, 87 (2001).

\bibitem{lukas} M.~Brandle and A.~Lukas,
Phys.\ Rev.\ D {\bf 65}, 064024 (2002).



\bibitem{Bardeen} J. M. Bardeen, Phys. Rev. D {\bf 22}, 1882 (1980).

\bibitem{MFB} V. F. Mukhanov, H. A. Feldman and R. H. Brandenberger,
        Phys. Rep. {\bf 215}, 203 (1992).

\bibitem{KS} H. Kodama and M. Sasaki, Prog. Theor. Phys. Suppl. {\bf
        78}, 1 (1984).





\bibitem{gordon} C.~Gordon, D.~Wands, B.~A.~Bassett and R.~Maartens,
Phys.\ Rev.\ D {\bf 63}, 023506 (2001).


\bibitem{StewartWalker}
J.M. Stewart and M. Walker, Proc. R. Soc. Lond. A {\bf 341}, 49 (1974).





\bibitem{bmr1} N.~Bartolo, S.~Matarrese and A.~Riotto,
Phys.\ Rev.\ D {\bf 64}, 083514 (2001).

\bibitem{bmr2} N.~Bartolo, S.~Matarrese and A.~Riotto,
Phys.\ Rev.\ D {\bf 64}, 123504 (2001).

\bibitem{bmr3} 
N.~Bartolo, S.~Matarrese and A.~Riotto,
hep-ph/0112261.


\bibitem{wbmr} D. Wands, N.~Bartolo, S.~Matarrese and A.~Riotto,
to appear.











\bibitem{book} A. Liddle and D. Lyth, {\it Cosmological Inflation and
Large
Scale Structure}, Cambridge University Press (2000).


\bibitem{pbb} G. Veneziano, Phys. Lett. {\bf B265}, 287 (1991); 
M. Gasperini and G. Veneziano, Astropart. Phys. {\bf 1}, 
  1 (1992); M. Gasperini and G. Veneziano, Phys. Rev.  {\bf D50}, 
  2519 (1994).

\bibitem{finelli}
F.~Finelli and R.~Brandenberger, hep-th/0112249.

\bibitem{cur1} K.~Enqvist and M.~S.~Sloth,
Nucl.\ Phys.\ B {\bf 626}, 395 (2002).

\bibitem{cur2}
D.~H.~Lyth and D.~Wands,
Phys.\ Lett.\ B {\bf 524}, 5 (2002);
T.~Moroi and T.~Takahashi,
hep-ph/0112335.

\bibitem{hanany} O.~J.~Ganor and A.~Hanany,
Nucl.\ Phys.\ B {\bf 474}, 122 (1996).



\bibitem{stro} A.~Strominger,
Phys.\ Lett.\ B {\bf 383}, 44 (1996).



\bibitem{us} A. Riotto and A. Zaffaroni, in preparation.

\end{thebibliography}
\end{document}